\documentclass[doublespacing]{elsart}
\usepackage{graphicx}
\usepackage{amssymb}
\usepackage{bm}

\begin{document}

\begin{frontmatter}

\title{Lancret helices}

\author{Alexandre F. da Fonseca,}
\author{C. P. Malta\corauthref{a1}}
\ead{coraci@if.usp.br}
\corauth[a1]{C. P. Malta}

\address{Instituto de F\'{\i}sica, Universidade de S\~ao Paulo,
USP\\ Rua do Mat\~ao, Travessa R 187, Cidade Universit\'aria,
05508-900, S\~ao Paulo, Brazil}

\begin{abstract}

Helical configurations of inhomogeneous symmetric rods with
non-constant bending and twisting stiffness are studied within the
framework of the Kirchhoff rod model. From the static Kirchhoff
equations, we obtain a set of differential equations for the {\it
curvature} and {\it torsion} of the centerline of the rod and the {\it
Lancret's theorem} is used to find helical solutions. We obtain a {\it
free standing} helical solution for an inhomogeneous rod whose
curvature and torsion depend on the form of variation of the {\it
bending coefficient} along the rod. These results are obtained for
inhomogeneous rods without intrinsic curvature, and for a particular
case of intrinsic curvature.

\end{abstract}

\end{frontmatter}


\section{Introduction}  

Helical filaments are tridimensional structures commonly found in
Nature. They can be seen in microscopic systems, as
biomolecu\-les~\cite{tamar}, bacterial fibers~\cite{wolge} and
nanosprings~\cite{mc}, and in macroscopic ones, as ropes, strings and
climbing plants~\cite{alain,tyler,pie}. Usually, the axis of all these
objects is modeled as a {\it circular helix}, {\it i. e.} a 3D-space
curve whose mathematical geometric properties, namely the {\it
curvature}, $k_F$, and the {\it torsion}, $\tau_F$, are
constant~\cite{nize,manfredo,dirk}. This kind of helical structure has
been shown to be a static solution of the Kirchhoff rod
model~\cite{nize}.

The Kirchhoff rod model~\cite{kirch,dill} has been proved to be a good
framework to study the statics~\cite{nize,vander,neuk} and
dynamics~\cite{goriely} of long, thin and inextensible elastic
rods. Applications of the Kirchhoff model range from
Biology~\cite{tamar,col1,tyler} to Engineering~\cite{sun} and,
recently, to Nanoscience~\cite{fonseca3}. In most cases, the rod or
filament is considered as being homogeneous, but the case of
nonhomogeneous rods have also been considered in the literature. It
has been shown that nonhomogeneous Kirchhoff rods may present spatial
chaos~\cite{holmes,davies}. In the case of planar rods, Domokos and
collaborators have provided some rigorous results for non-uniform
elasticae~\cite{domokos1} and for constrained Euler
buckling~\cite{domokos2,domokos3}. Deviations of the helical structure
of rods due to periodic variation of the Young's modulus were verified
numerically by da Fonseca, Malta and de Aguiar~\cite{fonseca0}.
Nonhomogeneous rods subject to given boundary conditions were studied
by da Fonseca and de Aguiar in~\cite{fonseca1}. The effects of a
nonhomogeneous mass distribution in the dynamics of unstable closed
rods have been analyzed by Fonseca and de Aguiar~\cite{fonseca2}.
Goriely and McMillen~\cite{alain2} studied the dynamics of cracking
whips~\cite{whip} and Kashimoto and Shiraishi~\cite{moto} studied
twisting waves in inhomogeneous rods.

The stability analysis of helical structures is of great importance in
the study of the elastic behavior of filamentary systems and has been
performed both experimentally~\cite{thompson} and
theoretically~\cite{goriely,tabor2,patrick}. It has been also shown
that the type of instability in twisted rods strongly depends on the
anisotropy of the cross section~\cite{thompson1,nize1,nadia}.

Here, we consider a rod with nonhomogeneous bending and twisting
coefficients varying along its arclength $s$, $B(s)$ and $C(s)$,
respectively. We are concerned with the following question: {\it is
there any helical solution for the stationary Kirchhoff equations in
the case of an inhomogeneous rod~?} The answer is `yes' and it will be
shown that the helical solution for an inhomogeneous rod with varying
bending coefficient cannot be the well known {\it circular helix}, for
which the curvature, $k_F$, and torsion, $\tau_F$, are constant. To
this purpose, we shall derive a set of differential equations for the
curvature and the torsion of the centerline of an inhomogeneous rod
and then apply the condition that a space curve must satisfy to be
helical: the {\it Lancret's theorem}. We shall obtain the simplest
helical solutions satisfying the {\it Lancret's theorem} and show that
they are {\it free standing helices}, {\it i.e.}, helices that are not
subjected to axial forces~\cite{tabor2}. A resulting helical structure
different from the {\it circular helix}, from now on, will be called a
{\it Lancret helix}.

According to the {\it fundamental theorem} for space
curves~\cite{dirk}, the curvature $k_F(s)$, and the torsion,
$\tau_F(s)$, completely determine a space curve, but for its position
in space. We shall show that the $k_F(s)$ and $\tau_F(s)$ of a {\it
Lancret helix} depend directly on the bending coefficient, $B(s)$, an
expected result since the centerline of the rod does not depend on the
twisting coefficient (see for example, Neukirch and
Henderson~\cite{neuk}).

Some motivations for this work are related to defects~\cite{kronert}
and distortions~\cite{geetha} in biological molecules. These defects
and distortions could be modeled as inhomogeneities along a continuous
elastic rod.

In Sec. II we review the general definition of a space curve, the {\it
Frenet} basis and the so-called {\it Lancret's theorem}. In Sec. III
we present the static Kirchhoff equations for an intrinsically
straight rod with varying stiffness, and derive the differential
equations for the curvature and torsion of the rod. In Sec. IV we use
the {\it Lancret's theorem} for obtaining helical solutions of the
static Kirchhoff equations and we show that they cannot be {\it
circular helices} if the bending coefficient is not constant. As
illustration, we compare a homogeneous rod with two simple cases of
inhomogeneous rods: (i) linear and (ii) periodic bending coefficient
varying along the rod. The {\it circular helix} has a well known
relation of the curvature and torsion with the radius and pitch of the
helix. In Sec. V we define a function involving all these variables in
such a way that for the {\it circular helix} its value is identically
null. We have verified, numerically, that this function approaches
zero for the inhomogeneous cases considered here. In Sec. VI we
analyse the cases of null torsion (straight and planar rods). Since
helical solutions of intrinsically straight rods are not dynamically
stable~\cite{tabor2}, in Sec. VII we consider a rod with a given
helical intrinsic curvature and we obtain, for this case, a helical
solution of the static Kirchhoff equations similar to that of an
intrinsically straight inhomogeneous rod. In Sec. VIII we summarize
the main results.

\section{Curves in space}

A curve in space can be considered as a path of a particle in motion.
The rectangular coordinates $(x,y,z)$ of the point on a curve can be
expressed as function of a parameter $u$ inside a given interval:
\begin{equation}
\label{xyz}
x=x(u) \; , \; \; y=y(u) \; , \; \; z=z(u) \; , \; \; \;
u_1\leq u\leq u_2 \; .
\end{equation}

We define the vector ${\mathbf{x}}(u) \equiv (x(u),y(u),z(u))$. If $u$
is the time, ${\mathbf{x}}(u)$ represents the trajectory of a
particle.

\subsection{The {\it Frenet} frame and the {\it Frenet-Serret} equations}

The vector tangent to the space curve at a given point $P$ is simply
$d{\mathbf{x}}/du$. It is possible to show \cite{dirk} that if the
arclength $s$ of the space curve is considered as its parameter, the
tangent vector at a given point $P$ of the curve ${\mathbf{x}}(s)$ is
a unitary vector. So, using the arclength $s$ to parametrize the
curve, we shall denote by ${\mathbf{t}}$ its tangent vector
\begin{equation}
\label{t}
{\mathbf{t}}=\frac{d{\mathbf{x}}}{ds} \; ,
\end{equation}
$\|{\mathbf{t}}\|=1$. The tangent vector ${\mathbf{t}}$ points in the
direction of increasing $s$.

The plane defined by the points $P_1$, $P_2$ and $P_3$ on the curve,
with $P_2$ and $P_3$ approaching $P_1$, is called the {\it osculating
plane} of the curve at $P_1$~\cite{dirk}. Given a point $P$ on the
curve, the principal normal at $P$ is the line, in the osculating
plane at $P$, that is perpendicular to the tangent vector at $P$. The
normal vector ${\mathbf{n}}$ is the unit vector associated to the
principal normal (its sense may be chosen arbitrarily, provided it is
continuous along the space curve).

From ${\mathbf{t}}.{\mathbf{t}}=1$, differentiating with respect to
$s$ (indicated by a prime) it follows that:
\begin{equation}
\label{tlinha}
{\mathbf{t}}.{\mathbf{t}}'=0 \; ,
\end{equation}
so that ${\mathbf{t}}$ and ${\mathbf{t}}'$ are orthogonal. It is
possible to show that ${\mathbf{t}}'$ lies in the osculating plane,
consequently ${\mathbf{t}}'$ is in the direction of
${\mathbf{n}}$. This allows us to write
\begin{equation}
\label{n}
{\mathbf{t}}'=k_F\,{\mathbf{n}} \; ,
\end{equation}
$k_F$ being called the {\it curvature} of the space curve at $P$.

The curvature measures the rate of change of the tangent vector when
moving along the curve. In order to measure the rate of change of the
osculating plane, we introduce the vector normal to this plane at $P$:
the {\it binormal unit vector} ${\mathbf{b}}$. At a point $P$ on the
curve, ${\mathbf{b}}$ is defined in such a way that
\begin{equation}
\label{v_b}
{\mathbf{b}}={\mathbf{t}}\times{\mathbf{n}} \; .
\end{equation}

The frame $\{{\mathbf{n}},{\mathbf{b}},{\mathbf{t}}\}$ can be taken as
a new frame of reference and forms the {\it moving trihedron} of the
curve. It is commonly called the {\it Frenet frame}. The rate of
change of the osculating plane is expressed by the vector
${\mathbf{b}}'$. It is possible to show that ${\mathbf{b}}'$ is
anti-parallel to the unit vector ${\mathbf{n}}$~\cite{dirk}. So we can
write
\begin{equation}
\label{torsion}
{\mathbf{b}}'=-\tau_F\,{\mathbf{n}} \; ,
\end{equation}
$\tau_F$ being called the {\it torsion} of the space curve at $P$.

The rate of variation of ${\mathbf{n}}$ \cite{dirk} can be obtained
straightforwardly. It is given by
\begin{equation}
\label{nlinha}
{\mathbf{n}}'=-k_F\,{\mathbf{t}}+\tau_F\,{\mathbf{b}} \; .
\end{equation}

The set of differential equations for
$\{{\mathbf{t}},{\mathbf{n}},{\mathbf{b}}\}$ is
\begin{equation}
\label{serret}
\begin{array}{l}
{\mathbf{t}}'=k_F\,{\mathbf{n}} \; , \\
{\mathbf{n}}'=-k_F\,{\mathbf{t}}+\tau_F\,{\mathbf{b}} \; , \\
{\mathbf{b}}'=-\tau_F\,{\mathbf{n}} \; ,
\end{array}
\end{equation}
and are known as the {\it formulas of Frenet} or the {\it
Serret-Frenet} equations~\cite{dirk}.

\subsection{The {\it Fundamental theorem} of space curves}

A space curve parametrized by its arclength $s$ is defined by a
vectorial function ${\mathbf{x}}(s)$. The form of ${\mathbf{x}}(s)$
depends on the choice of the coordinate system. Nevertheless, there
exists a form of characterization of a space curve given by a relation
that is independent of the coordinates. This relation gives the {\it
natural equation} for the curve.

$k_F(s)$ gives the natural equation in the case of planar
curves. Indeed, if $\varphi$ is the angle between the tangent vector
of the planar curve and the $x$-axis of the coordinate system, it is
possible to show that $k_F=d\varphi/ds$. Since $\cos(\varphi)=dx/ds$
and $\sin(\varphi)=dy/ds$, knowing $k_F(s)$, then $\varphi(s)$, $x(s)$,
and $y(s)$ of the planar curve can be obtained immediatly:
\begin{equation}
\label{xyfi}
\varphi(s)=\int_{s_0}^{s}k_F(s)\,ds  , \; \;
x(s)=\int_{s_0}^{s}\cos\varphi (s)ds  , \; \;
y(s)=\int_{s_0}^{s}\sin\varphi (s)ds  . 
\end{equation}

In the case of non-planar curves, if we have {\it two single valued
continuous functions} $k_F(s)$ {\it and } $\tau_F(s),\;s>0,$ {\it then
there exists one and only one space curve, determined but for its
position in space, for which} $s$ {\it is the arclength, } $k_F(s)$
{\it the curvature, and} $\tau_F(s)$ {\it the torsion.} It is the {\it
Fundamental theorem} for space curves \cite{dirk}. The functions
$k_F(s)$ and $\tau_F(s)$ provide the {\it natural equations} of the
space curve.

\subsection{Curves of constant slope: the {\it Lancret's theorem}}

A space curve ${\mathbf{x}}(s)$ is a {\it helix} if the lines tangent
to ${\mathbf{x}}$ make a constant angle with a fixed direction in
space (the helical axis)~\cite{manfredo,dirk}. Denoting by
${\mathbf{a}}$ the unit vector of this direction, a helix satisfies
\begin{equation}
\label{helix}
{\mathbf{t}}.{\mathbf{a}}=\cos\alpha= 
\mbox{constant} \; .
\end{equation}
Differentiating Eq. (\ref{helix}) with respect to $s$ gives
${\mathbf{a}}.{\mathbf{n}}=0$. Therefore ${\mathbf{a}}$ lies in the
plane determined by the vectors ${\mathbf{t}}$ and ${\mathbf{b}}$:
\begin{equation}
\label{v_a}
{\mathbf{a}}={\mathbf{t}}\cos\alpha+{\mathbf{b}}\sin\alpha \; .
\end{equation}
Differentiating Eq. (\ref{v_a}) with respect to $s$, gives
\[
0=(k_F\,\cos\alpha-\tau_F\,\sin\alpha){\mathbf{n}}\; ,
\]
or
\begin{equation}
\label{lancret}
\frac{k_F}{\tau_F}=\tan\alpha=\mbox{constant.}
\end{equation}

This result says that {\it for curves of constant slope the ratio of
curvature over torsion is constant}. Conversely, given a regular curve
for which the equation (\ref{lancret}) is satisfied, it is possible to
find \cite{dirk} a constant angle $\alpha$ such that
\[
{\mathbf{n}}\left(k_F\cos\alpha-\tau_F\sin\alpha\right)=0 \; ,
\]
\[ 
\frac{d}{ds}\left({\mathbf{t}}\cos\alpha+{\mathbf{b}}\sin\alpha\right)
=0 \; ,
\]
implying that the vector
$\mathbf{a}={\mathbf{t}}\cos\alpha+{\mathbf{b}}\sin\alpha$ is the unit
vector along the axis. Moreover, ${\mathbf{a}}.{\mathbf{t}}=
\cos\alpha$=constant, so that the curve has constant slope. This
result can be expressed as:

{\it A necessary and suficient condition for a space curve to be a
curve of constant slope (a helix) is that the ratio of curvature over
torsion be constant}. It is the well known {\it Lancret's theorem},
dated of 1802 and first proved by B. de Saint
Venant~\cite{dirk,venant}.

If a helical curve ${\mathbf{x}}(s)$ is projected onto the plane
perpendicular to ${\mathbf{a}}$, the vector ${\mathbf{x}}_1(s)$ 
representing this projection is given by
\begin{equation}
\label{x1}
{\mathbf{x}}_1(s) ={\mathbf{x}}-({\mathbf{x}}.{\mathbf{a}}){\mathbf{a}} \; .
\end{equation}
It is possible to show \cite{dirk} that the curvature $k_1$ of the
projected curve is given by:
\begin{equation}
\label{k1}
k_1(s)=\frac{k_F(s)}{\sin^{2}\alpha} \; . 
\end{equation}

The shape of the planar curve obtained by projecting a helical curve
onto the plane perpendicular to its axis is used to characterize
it. For example, the well known {\it circular helix} projects a circle
onto the plane perpendicular to its axis. The {\it spherical helix}
projects an arc of an epicycloid onto a plane perpendicular to its
axis~\cite{dirk}. The logarithmic spiral is the projection of a
helical curve called {\it conical helix}~\cite{dirk}.

\section{The static Kirchhoff equations} 

The statics and dynamics of long and thin elastic rods are governed by
the Kirchhoff rod model. In this model, the rod is divided in segments
of infinitesimal thickness to which the Newton's second law for the
linear and angular momentum are applied. We derive a set of partial
differential equations for the averaged forces and torques on each
cross section and for a triad of vectors describing the shape of the
rod. The set of PDE are completed with a linear constitutive relation
between torque and twist.

The central axis of the rod, hereafter called centerline, is
represented by a space curve ${\mathbf{x}}$ parametrized by the
arclength $s$. A {\it Frenet} frame is defined for this space curve as
described in the previous section. For a physical filament the use of
a local basis, $\{{\mathbf{d}}_1,{\mathbf{d}}_2,{\mathbf{d}}_3\}$, to
describe the rod has the advantage of taking into account the twist
deformation of the filament. This local basis is defined such that
${\mathbf{d}}_3$ is the vector tangent to the centerline of the rod
(${\mathbf{d}}_3={\mathbf{t}}$), and ${\mathbf{d}}_1$ and
${\mathbf{d}}_2$ lie on the cross section plane. The local basis is
related to the {\it Frenet} frame
$\{{\mathbf{n}},{\mathbf{b}},{\mathbf{t}}\}$ through
\begin{equation}
\label{def_xi}
({\mathbf{d}}_1 \; {\mathbf{d}}_2 \; {\mathbf{d}}_3)=
({\mathbf{n}} \; {\mathbf{b}} \; {\mathbf{t}})
\left(
\begin{array}{ccc}
\cos\xi & -\sin\xi & 0  \\
\sin\xi & \cos\xi & 0  \\
0 & 0 & 1
\end{array}
\right)  \; ,
\end{equation}
where the angle $\xi$ is the amount of twisting of the local basis
with respect to ${\mathbf{t}}$.

In this paper, we are concerned with equilibrium solutions of the
Kirchhoff model, so our study departs from the static Kirchhoff
equations~\cite{foot2}. In scaled variables, for intrinsically
straight isotropic rods, these equations are:
%
\begin{eqnarray}
&&\mathbf{F}'=0 \; , \label{kir1_a} \\
&&\mathbf{M}'=\mathbf{F}\times\mathbf{d}_{3} \; , \label{kir1_b} \\
&&\mathbf{M}=B(s)\,k_1\,\mathbf{d}_1+B(s)\,k_2\,\mathbf{d}_2+
C(s)\,k_3\,\mathbf{d}_3 \; , \label{kir1_c}
\end{eqnarray}
%
the vectors $\mathbf{F}$ and $\mathbf{M}$ being the resultant force,
and corresponding moment with respect to the centerline of the rod,
respectively, at a given cross section. As in the previous section,
$s$ is the arclength of the rod and the prime $'$ denotes
differentiation with respect to $s$. $k_i$ are the components of the
twist vector, $\mathbf{k}$, that controls the variations of the
director basis along the rod through the relation
\begin{equation}
\label{dis}
{\mathbf{d}}'_i={\mathbf{k}}\times{\mathbf{d}}_i \; , \; \; i=1,2,3 \; . 
\end{equation}
$k_1$ and $k_2$ are related to the curvature of the centerline of the
rod $(k_F=\sqrt{k^{2}_{1}+k^{2}_{2}})$ and $k_3$ is the twist
density. $B(s)$ and $C(s)$ are the bending and twisting coefficients
of the rod, respectively. In the case of macroscopic filaments the
bending and twisting coefficients can be related to the cross section
radius and the Young's and shear moduli of the rod. Writing the force
$\mathbf{F}$ in the director basis,
\begin{equation}
\label{F}
{\bf{F}}=f_1{\bf{d}}_1+f_2{\bf{d}}_2+f_3{\bf{d}}_3 \; , 
\end{equation}
the equations (\ref{kir1_a}--\ref{kir1_c}) give the following
differential equations for the components of the force and twist
vector:
%
\begin{eqnarray}
f'_1-f_2\, k_3+f_3\, k_2=0 \; , \label{a}  \\
f'_2+f_1\, k_3-f_3\, k_1=0 \; , \label{b}  \\
f'_3-f_1\, k_2+f_2\, k_1=0 \; , \label{c}  \\
(B(s)\, k_1)'+(C(s)-B(s))\, k_2\, k_3-f_2=0 \; , \label{d} \\
(B(s)\, k_2)'-(C(s)-B(s))\, k_1\, k_3+f_1=0 \; , \label{e} \\
(C(s)\ k_3)'=0 \; . \label{f} 
\end{eqnarray}
%
The equation (\ref{f}) shows that the component $M_3=C(s)\,k_3$ of the
moment in the director basis (also called {\it torsional moment}), is
constant along the rod, consequently the twist density $k_3$ is
inversely proportional to the twisting coefficient $C(s)$
\begin{equation}
\label{k3}
k_3(s)=\frac{M_3}{C(s)} \; .
\end{equation}

In order to look for helical solutions of the Eqs. (\ref{a}--\ref{f})
the components of the twist vector $\mathbf{k}$ are expressed as
follows:
%
\begin{eqnarray}
k_1&=&k_F(s)\sin\xi \; , \label{hel_k1} \\
k_2&=&k_F(s)\cos\xi \; , \label{hel_k2} \\
k_3&=&\xi'+\tau_F(s) \; , \label{hel_k3}
\end{eqnarray}
%
where $k_F(s)$ and $\tau_F(s)$ are the curvature and torsion,
respectively, of the space curve that defines the centerline of the
rod and $\xi$ is given by Eq. (\ref{def_xi}). If the rod is
homogeneous, the helical solution has constant $k_F$ and $\tau_F$, and
$\xi'$ is proved to be null~\cite{tyler}.

Substituting Eqs. (\ref{hel_k1}--\ref{hel_k3}) in
Eqs. (\ref{a}--\ref{f}), extracting $f_1$ and $f_2$ from Eqs. (\ref{e})
and (\ref{d}), respectively, differentiating them with respect to $s$,
and substituting in Eqs. (\ref{a}), (\ref{b}) and (\ref{c}), gives the
following set of nonlinear differential equations:
%
\begin{eqnarray}
[M_3\,k_F(s)-B(s)\,k_F(s)\,\tau_F(s)]'-(B(s)\,k_F(s))'\,\tau_F(s)
=0 \; ,  \label{dif_1} \\
(B(s)\,k_F(s))''+k_F(s)\,\tau_F(s)[M_3-B(s)\,\tau_F(s)]
-f_3(s)\,k_F(s)=0 \; , \label{dif_2}  \\
(B(s)\,k_F(s))'\,k_F(s)+f'_3(s)=0 \; . \label{dif_3}  
\end{eqnarray}
%
Appendix A presents the details of the derivation of
Eqs. (\ref{dif_1}--\ref{dif_3}).

The Eqs. (\ref{dif_1}--\ref{dif_3}) for the curvature, $k_F$, and
torsion, $\tau_F$ do not depend on the twisting coefficient,
$C(s)$. Therefore, the centerline of an inhomogeneous rod does not
depend on the twisting coefficient like in the case of homogeneous
rods (see, for example, Eqs. (13) and (14) of Ref. \cite{neuk}).

Langer and Singer~\cite{langer} have obtained a set of first-order
ordinary differential equations for the curvature and torsion of the
centerline of a homogeneous rod that contains terms proportional to
$k_F^2$ and $\tau^2_F$. The Eqs. (\ref{dif_1}--\ref{dif_3}) have the
advantage of involving only terms linear in $k_F$ and $\tau_F$.

\section{Helical solutions of inhomogeneous rods}

In order to find helical solutions for the static Kirchhoff equations,
we apply the {\it Lancret's theorem} to the general equations
(\ref{dif_1}--\ref{dif_3}). We first rewrite the {\it Lancret's
theorem} in the form:
\begin{equation}
\label{lan2}
k_F(s)=\beta\,\tau_F(s) \; , 
\end{equation}
with $\beta\neq0$. From Eq. (\ref{lancret}),
\begin{equation}
\label{beta}
\beta\equiv\tan\alpha= \mbox{Constant} \; .
\end{equation}
Substituting Eq. (\ref{lan2}) in Eq. (\ref{dif_1}) we obtain
\begin{equation}
\label{eq_B1}
\tau_F'\,(M_3-B\,\tau_F)-2\,\tau_F\,(B\,\tau_F)'=0 \; .
\end{equation}
Substituting Eq. (\ref{lan2}) in Eq. (\ref{dif_2}) and extracting
$f_3$, we obtain
\begin{equation}
\label{f3_2}
f_3=\frac{(B\,\tau_F)''}{\tau_F}+\tau_F\,(M_3-B\,\tau_F) \; .
\end{equation}
Differentiating $f_3$ with respect to $s$ and substituting in
Eq. (\ref{dif_3}) we obtain the following differential equation for
$\tau_F$:
\begin{equation}
\label{eq_B2}
\frac{(B\,\tau_F)'''}{\tau_F}-\frac{(B\,\tau_F)''\tau_F'}{\tau_F^2}+
(\beta^2+1)\,\tau_F\,(B\,\tau_F)'=0 \; ,
\end{equation}
where the Eq. (\ref{eq_B1}) was used to simplify the above
equation. One immediate solution for this differential equation is
\begin{equation}
\label{sol1}
(B\,\tau_F)'=0 \; ,
\end{equation}
that substituted in Eq. (\ref{eq_B1}) gives
\begin{equation}
\label{aux5}
\tau_F'\,(M_3-B\tau_F)=0 \; .
\end{equation}
For non-constant $\tau_F$, the Eq. (\ref{aux5}) gives the following
solution for $\tau_F$:
\begin{equation}
\label{tf_1}
\tau_F(s)=\frac{M_3}{B(s)} \; .
\end{equation}
Substituting the Eqs. (\ref{sol1}) and (\ref{tf_1}) in
Eq. (\ref{f3_2}) we obtain that
\begin{equation}
\label{f3Zero}
f_3(s)=0 \; .
\end{equation}
Substituting Eq. (\ref{tf_1}) in (\ref{lan2}) we obtain:
\begin{equation}
\label{kf_1}
k_F(s)=\beta\frac{M_3}{B(s)}  \; .
\end{equation}

Substituting Eq. (\ref{def_xi}) in Eq. (\ref{F}), the force ${\bf{F}}$
becomes
\begin{equation}
\label{F_frenet}
{\bf{F}}=(f_1\cos\xi-f_2\sin\xi)\,{\bf{n}}+
(f_1\sin\xi+f_2\cos\xi)\,{\bf{b}}+f_3\,{\bf{t}} \; ,
\end{equation}
where $\{{\bf{n}},{\bf{b}},{\bf{t}}\}$ is the Frenet basis. Using the
Eqs. (\ref{f1r}) and (\ref{f2r}) for $f_1$ and $f_2$ (Appendix A), we
obtain
\begin{equation}
\label{F2}
{\bf{F}}=-(B\,k_F)'\,{\bf{n}}+k_F\,[M_3-B\,\tau_F]\,{\bf{b}}+
f_3\,{\bf{t}} \; ,
\end{equation}
where $f_3$, in the inhomogeneous case, must satisfy the
Eq. (\ref{dif_3}).

Substituting the Eqs. (\ref{tf_1}--\ref{kf_1}) in the Eq. (\ref{F2}),
and using Eq. (\ref{beta}), it follows ${\bf{F}}=0$. Therefore, the
helical solutions satisfying (\ref{sol1}) are {\it free standing}.

Now, we prove that a {\it circular helix} cannot be a solution of the
static Kirchhoff equations for a rod with varying bending
stiffness. If a helix is circular, $k_F'=0$ and $\tau_F'=0$, and from
Eq. (\ref{dif_1}) we obtain:
\begin{equation}
\label{Blinha}
2\,k_F\,\tau_F\,B'=0 \; .
\end{equation}
Since $B'(s)\neq0$, Eq. (\ref{Blinha}) will be satisfied only if $k_F=0$
and/or $\tau_F=0$. Therefore, it is not possible to have a {\it
circular helix} as a solution for a rod with varying bending
coefficient.

The solutions for the curvature $k_F$, Eq. (\ref{kf_1}), and the
torsion $\tau_F$, Eq. (\ref{tf_1}), can be used to obtain the unit
vectors of the {\it Frenet} frame (by integration of the
Eqs. (\ref{serret})). From Eqs. (\ref{kf_1}), (\ref{beta}) and
(\ref{v_a}), we can obtain $\alpha$ and ${\bf{a}}$ once $k_F(0)$,
$M_3$ and $B(s)$ are given. By choosing the $z$-direction of the fixed
cartesian basis as the direction of the unit vector ${\mathbf{a}}$, we
can integrate ${\mathbf{t}}$ in order to obtain the three-dimensional
configuration of the centerline of the rod.

\begin{figure}[ht]
  \begin{center}
  \includegraphics[height=44mm,width=19mm,clip]{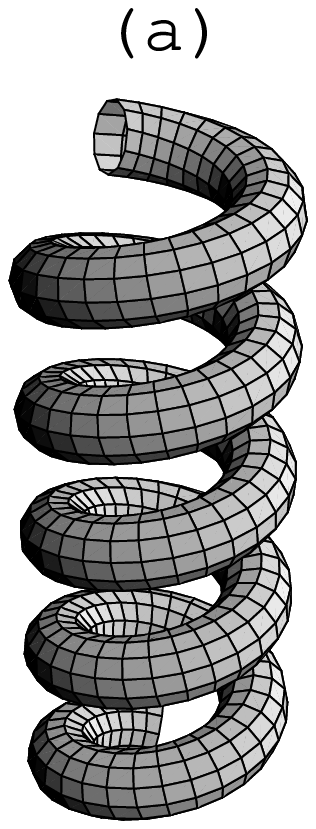}
  \includegraphics[height=44mm,width=26mm,clip]{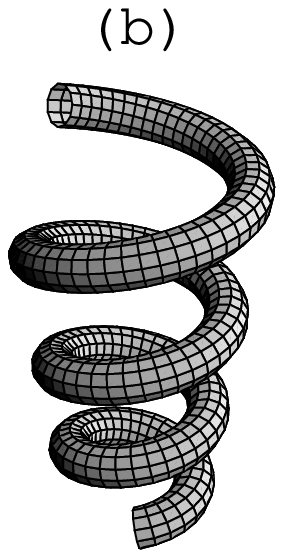}
  \includegraphics[height=44mm,width=23mm,clip]{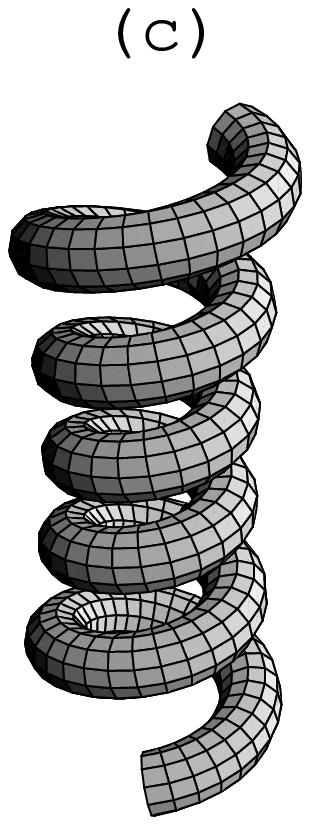}
  \end{center} 
  \caption{Helical solutions of the Kirchhoff equations using the {\it
  Lancret's Theorem}. (a) circular helix solution for an inhomogeneous
  rod with constant bending coefficient $B_a=1$ (\ref{r1}); (b) and (c)  
  Lancret helices for inhomogeneous rod with bending coefficient
  given by Eqs. (\ref{r2}) and (\ref{r3}), respectively. The
  parameters, in scaled units, are $M_3=0.05$, $\Gamma=0.9$,
  and the total length of the rod is $L=130$. $k_F(0)=0.24$ for the helical 
  solutions displayed in panels (a) and (b), and $k_F(0)=0.22$ for panel 
  (c). }  
\label{fig1}
\end{figure}

Figure~\ref{fig1} displays the helical solution of the static Kirchhoff
equations for rods with bending coefficients given by
%
\begin{eqnarray}
\mbox{Fig \ref{fig1}a:}\;\;\;B_{a}(s)&=&1 \; , \label{r1} \\
\mbox{Fig \ref{fig1}b:}\;\;\;B_{b}(s)&=&1+0.007\,s \; \; , \label{r2} \\
\mbox{Fig \ref{fig1}c:}\;\;\;B_{c}(s)&=&1+0.1\,\sin(0.04s+2) \; \; . \label{r3}
\end{eqnarray}
%
The case of constant bending (\ref{r1}) produces the well known {\it
circular helix} displayed in
Fig. \ref{fig1}a. Figs. \ref{fig1}b--\ref{fig1}c show that
non-constant bending coefficients (Eqs. (\ref{r2}--\ref{r3})) do not
produce a circular helix.

The helical solutions displayed in Fig. \ref{fig1} satisfy the {\it
Lancret's theorem}, Eq. (\ref{lancret}). The tridimensional helical
configurations displayed in Fig. \ref{fig1} were obtained by
integrating the {\it Frenet-Serret} equations (\ref{serret}) using the
following initial conditions for the Frenet frame:
${\mathbf{t}}(s=0)=(0,\sin\alpha,\cos\alpha)$,
${\mathbf{n}}(s=0)=$~$(-1,0,0)$ and
${\mathbf{b}}(s=0)=$~$(0,-\cos\alpha,\sin\alpha)$. This choice ensures
that the $z$-axis is parallel to the direction of the helical axis,
vector ${\mathbf{a}}$. The centerline of the helical rod is a space
curve ${\mathbf{x}}(s)=(x(s),y(s),z(s))$ that is obtained by
integration of the tangent vector ${\mathbf{t}}(s)$. We have taken the
helical axis as the $z$-axis and placed the initial position of the
rod at $x(0)=1/k_1(0)$, $y(0)=0$ and $z(0)=0$ (in scaled units), where
$k_1(0)$ is the curvature of the planar curve at $s=0$ obtained by
projecting the space curve onto the plane perpendicular to the helical
axis (Eq. (\ref{k1})). From Eq. (\ref{k1}) we have
\begin{equation}
\label{k10}
k_1(0)=\frac{k_F(0)}{\sin^2\alpha} \; . 
\end{equation}
Using the Eq. (\ref{beta}), it follows that
\begin{equation}
\label{sinalfa}
\sin^2\alpha=\frac{\beta^2}{1+\beta^2}\; .
\end{equation}
From Eq. (\ref{kf_1}), setting $s=0$, we get
\begin{equation}
\label{beta2}
\beta=\frac{k_F(0)\,B(0)}{M_3} \; .
\end{equation}
Substituting Eqs. (\ref{sinalfa}) and (\ref{beta2}) in
Eq. (\ref{k10}), we obtain:
\begin{equation}
\label{X0}
x(0)=\frac{1}{k_1(0)}=\frac{k_F(0)\,B^2(0)}{M_3^2+k_F^2(0)B^2(0)} \; .
\end{equation}
$k_F(0)$ and $M_3$ are free parameters that have been chosen so that
the helical solutions displayed in Fig. \ref{fig1} have the same angle
$\alpha$. The parameters $k_F(0)=0.24$ and $M_3=0.05$ give
$x(0)\simeq4$ for the helical solutions displayed in Figs. \ref{fig1}a
and \ref{fig1}b, and the parameters $k_F(0)=0.22$ and $M_3=0.05$ give
$x(0)\simeq4.36$ for the helical solution displayed in the
Fig. \ref{fig1}c.

For short the projection of the space curve onto the plane
perpendicular to the helical axis will be called {\it projected
curve}. As mentioned in Sec. II, the circle is the projected curve of
the most common type of helix, the {\it circular helix}.

\begin{figure}[ht]
  \begin{center}
  \includegraphics[height=23mm,width=23mm,clip]{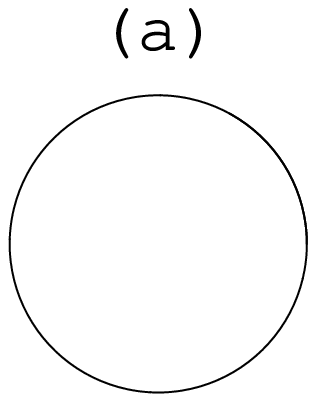}
  \includegraphics[height=23mm,width=23mm,clip]{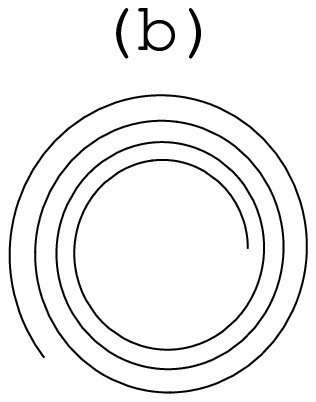}
  \includegraphics[height=23mm,width=23mm,clip]{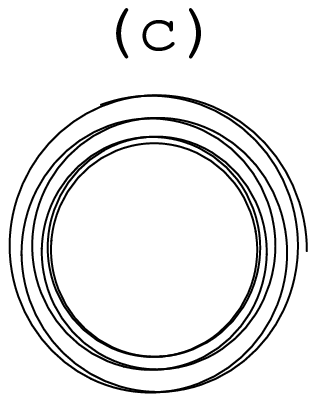}
  \end{center} \caption{(a), (b) and (c) are projected curves of the
  helical solutions displayed in Fig. \ref{fig1}a, Fig. \ref{fig1}b
  and Fig. \ref{fig1}c, respectively. We used Eq. (\ref{x1}) to obtain
  the projected curves.}
\label{fig2}
\end{figure}

Fig. \ref{fig2} displays the projected curves related to the helical
solutions displayed in Fig. \ref{fig1}. Fig. \ref{fig2}a shows that
the helical solution of the inhomogeneous rod with constant bending
coefficient projects a circle onto the plane perpendicular to the
helical axis.

If required, the natural equations for the projected curves displayed
in Fig. \ref{fig2} are easily obtained, for instance, by substitution
of the solution (Eq. (\ref{kf_1})) for the curvature $k_F(s)$ of the
helical rod into Eq. (\ref{k1}). The natural equation of the projected
curve is given by its curvature,
\begin{equation}
\label{natural}
k_{1}(s)=\frac{\beta M_3}{B(s)}\sin^{-2}\alpha \; ,
\end{equation}
where $\beta$ and $\sin^{-2}\alpha$ can be obtained by
Eqs. (\ref{sinalfa}) and (\ref{beta2}). Then, in the
Eq.(\ref{natural}), setting $B(s)=B_i(s)$, $i=a,b,c$, as given in
Eqs.~(\ref{r1}--\ref{r3}), produces the natural equation for the
corresponding projected curve displayed in Fig.~\ref{fig2}. The
helical rod displayed in Fig. \ref{fig1}b is a {\it conical helix}
since the radius of curvature of its projected curve (inverse of
$k_1(s)$) is a {\it logarithmic spiral} ($1/k_1(s)$ is a linear
function of $s$~\cite{dirk}).

From Eqs. (\ref{hel_k1}--\ref{hel_k3}), (\ref{k3}) and (\ref{tf_1}) we
obtain the variation of the angle $\xi$ between the local basis,
${\mathbf{d}}_i$, $i=1,2,3$, and the {\it Frenet} frame,
$\{{\mathbf{n}},{\mathbf{b}},{\mathbf{t}}\}$:
\begin{equation}
\label{xilinha}
\xi'=k_3(s)-\tau_F(s)=\left( \frac{M_3}{C(s)} -
\frac{M_3}{B(s)} \right)=M_3\frac{B(s)-C(s)}{B(s)\,C(s)}\; .
\end{equation}
Eq. (\ref{xilinha}) shows that $\xi'\ne 0$ for the general case of
$B(s)\ne C(s)$, {\it i. e.} helical filaments corresponding to
inhomogeneous rods {\it are not twistless}. The {\it circular helix}
is a helicoidal solution for the centerline of an inhomogeneous rod
having constant bending coefficient. We emphasize that the
inhomogeneous rod is not twistless in contrast with the homogeneous
case where it has been proved that $\xi'=0$~\cite{tyler}. 

A homogeneous rod has $B(s)$ and $C(s)$ constant so that $k_3=$
Constant (from Eq. (\ref{f})). Since $\xi'$ has been proved to be null
for a helical solution of a homogeneous rod (see
reference~\cite{tyler}), Eq. (\ref{hel_k3}) shows that the torsion
$\tau_F$ must be a constant. In order to satisfy the {\it Lancret's
theorem} (Eq. (\ref{lancret})) the curvature $k_F$ of the helical
solution must also be a constant. Therefore, the only type of helical
solution for a homogeneous rod is the circular helix, while an
inhomogeneous rod may present other types of helical structures.

\section{Radius and Pitch of the helical solution}

The {\it radius} $\mathcal{R}$ of a helix is defined as being the
distance of the space curve to its axis. The {\it pitch} $\mathcal{P}$
of a helix is defined as the height of one helical turn, {\it i.e.},
the distance along the helical axis of the initial and final points of
one helical turn.

For a circular helix, $\mathcal{R}$, $\mathcal{P}$, $k_F$ and $\tau_F$
are constant, and it is easy to prove that
\begin{equation}
\label{lambda}
\lambda=\left(\sqrt{\mathcal{R}^2+\mathcal{P}^2/(4\pi^2)}\right)^{-1}=
\sqrt{k_{F}^2+\tau_{F}^2} \; .
\end{equation}

For other types of helix, it constitutes a very hard problem in
differential geometry to obtain the relation between the curvature
$k_F$ and the torsion $\tau_F$ with the radius $\mathcal{R}$ and the
pitch $\mathcal{P}$. We have seen in Sec. II that the definitions of
curvature and torsion involve the calculation of the modulus of the
tangent and normal vectors derivative with respect to the arclength of
the rod. We also saw that the {\it Frenet-Serret} differential
equations for the {\it Frenet} frame depend on the curvature and
torsion. The difficulty of integration of the {\it Frenet-Serret}
equations for the general case where $k_F(s)$ and $\tau_F(s)$ are
general functions of $s$ poses the problem of finding an analytical
solution for the centerline of the rod, thus the difficulty of
relating non constant curvature and torsion with non constant radius
and pitch. Due to this difficulty we shall test the possibility of
generalizing the relation (\ref{lambda}) to the present inhomogeneous
case. In order to do so, from the equation (\ref{lambda}) we define:
\begin{equation}
\label{Glamb}
g_{\lambda}(s)\equiv\left(\sqrt{\mathcal{R}^{2}(s)+\mathcal{P}^{2}(s)/
(4\pi^2)}\right)^{-1}
- \sqrt{k_{F}^{2}(s)+\tau_{F}^{2}(s)} \; ,
\end{equation}
where $\mathcal{R}(s)$ and $\mathcal{P}(s)$ are the radius and the
pitch of the helical structure as function of $s$. In the case of a
circular helix, from Eq. (\ref{lambda}), $g_{\lambda}(s)=0$ for all
$s$.

Since the $z$-axis is defined as being the axis of the helical
solution we can calculate the radius $\mathcal{R}(s)$ through:
$\mathcal{R}(s)=\sqrt{x^{2}(s)+y^{2}(s)}$, where $x(s)$ and $y(s)$ are
the $x$ and $y$ components of the vector position of the centerline of
the helical rod. 

The pitch of the helix is the difference between the $z$-coordinate of
the initial and final positions of one helical turn. A helical turn
can be defined such that the projection of the vector position of the
spatial curve along the $xy$-plane (vector ${\mathbf{x}}_1$ of
Eq. (\ref{x1})), rotates of $2\pi$ around the $z$-axis.

\begin{figure}[ht]
  \begin{center}
  \includegraphics[height=40mm,width=70mm,clip]{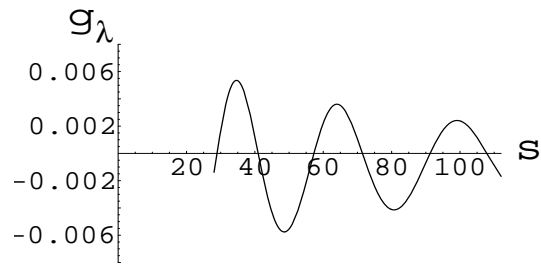} 
  \end{center}
  \caption{$g_{\lambda}(s)$ for the free standing helix solution
  displayed in Fig. \ref{fig1}b.}
\label{fig3}
\end{figure}

Fig. \ref{fig3} shows $g_{\lambda}(s)$ for the free standing helix of
Fig. \ref{fig1}b. We see that $g_{\lambda}(s)$ oscillates, its maximum
amplitude being smaller than $0.006$. For the helical shape displayed
in Fig. \ref{fig1}c we found that the maximum value of
$g_{\lambda}(s)$ is smaller than $0.008$ (data not shown). While for a
circular helix $g_{\lambda}=0$, for the free standing helices
displayed in Fig. \ref{fig1}b and Fig. \ref{fig1}c the function
$g_{\lambda}$ oscillates around zero with small amplitude. 

The small amplitude of these oscillations suggests that the relations
$\mathcal{R}(s)\simeq k_F(s)[\sqrt{k_{F}^2(s)+
\tau_{F}^2(s)}]^{-1}$ and $\mathcal{P}(s)\simeq 2\pi \tau_F(s)
[\sqrt{k_{F}^2(s)+\tau_{F}^2(s)}]^{-1}$, valid for circular helices,
could be used to derive approximate functions for the radius and the
pitch of different types of helical structures, but the oscillatory
behavior indicates that these relations are not simple functions of
the geometric features of the helix.

\section{Straight and planar inhomogeneous rods}

Straight rods ($k_F=0$), and planar rods ($k_F\neq0$), have null
torsion ($\tau_F=0$), and constitute particular cases of helices. In
both cases there is at least one direction in  space that makes a
constant angle $\alpha=\pi/2$ with the vector tangent to the rod centerline.

The straight inhomogeneous rod is a solution of the static Kirchhoff
equations that has non-constant twist density (Eq. (\ref{k3})), in
contrast with the homogeneous case for which the twist density is
constant. 

The twisted planar ring ($k_F=$ Constant) is a solution of the static
Kirchhoff equations only if the bending coefficient can be written in the
form:
\begin{equation}
\label{ring}
B(s)=A_0\cos(k_F\,s)+B_0\sin(k_F\,s)+ C_I/k_F^{2} \; ,
\end{equation}
with $A_0$, $B_0$ and $C_I$ constant. If $k_F$ is function of $s$
(instead of being a constant) there exist no solutions for
Eqs. (\ref{a}--\ref{f}). So, the existence of a planar solution
related to the general form of the components of the twist vector
given by equations (\ref{hel_k1}--\ref{hel_k3}) requires $k_F=$
Constant.

\section{Helical structure with intrinsic curvature}

The helical shape displayed in Fig.~\ref{fig1}b resembles that
exhibited by the tendrils of some climbing plants. In these plants the
younger parts have smaller cross section diameter, giving rise to
non-constant bending coefficient. The main difference between the
solution displayed in Fig.~\ref{fig1}b and the tendrils of climbing
plants is that the solution in Fig.~\ref{fig1}b was obtained for an
intrinsically straight rod while the tendrils have intrinsic curvature
~\cite{tyler}.

The tendrils of climbing plants are stable structures while the
helical solution displayed in Fig.~\ref{fig1}b is not stable because
the rod is intrinsically straight~\cite{tabor2}. We shall show that a
rod with intrinsic curvature may have a static solution of the
Kirchhoff equations similar to that displayed in Fig.~\ref{fig1}b.

The intrinsic curvature of a rod is introduced in the Kirchhoff model
through the components of the twist vector, $\mathbf{k}^{(0)}$, in the
{\it unstressed configuration} of the rod as
%
\begin{eqnarray}
k^{(0)}_1&=&k_F^{(0)}(s)\sin\xi \; , \label{ks0_1} \\
k^{(0)}_2&=&k_F^{(0)}(s)\cos\xi \; , \label{ks0_2} \\
k^{(0)}_3&=&\xi'+\tau_F^{(0)}(s) \; , \label{ks0_3}
\end{eqnarray}
%
where $k_F^{(0)}(s)$ and $\tau_F^{(0)}(s)$ are the curvature and
torsion of the space curve that represents the axis of the rod in its
unstressed configuration, simply called {\it intrinsic curvature} of
the rod. We consider that the unstressed configuration of the axis of
the rod forms a helical space curve with the intrinsic curvature
satisfying
%
\begin{eqnarray}
&&B(s)\,k_F^{(0)}(s)=K_0 \; , \label{int_a} \\
&&B(s)\,\tau_F^{(0)}(s)=T_0 \; , \label{int_b}
\end{eqnarray}
%
where $K_0$ and $T_0$ are constant and $B(s)$ is the bending
coefficient of the rod.

The linear constitutive relation (Eq. (\ref{kir1_c})) becomes
\begin{equation}
\label{lincons}
{\mathbf{M}}=B(s)(k_1-k_1^{(0)}){\mathbf{d}}_1 +
B(s)(k_2-k_2^{(0)}){\mathbf{d}}_2 + C(s)(k_3-k_3^{(0)}){\mathbf{d}}_3 
\; ,
\end{equation}
where $C(s)$ is the twisting coefficient of the rod. The static
Kirchhoff equations for this case, Eqs. (\ref{kir1_a}), (\ref{kir1_b})
and Eq. (\ref{lincons}), are given by
%
\begin{eqnarray}
f'_1-f_2\, k_3+f_3\, k_2=0 \; , \label{ica}  \\
f'_2+f_1\, k_3-f_3\, k_1=0 \; , \label{icb}  \\
f'_3-f_1\, k_2+f_2\, k_1=0 \; , \label{icc}  \\
(B(s)(k_1-k_1^{(0)}))'-B(s)(k_2-k_2^{(0)})k_3+C(s)(k_3-k_3^{(0)})-f_2=0 
\; , \label{icd} \\
(B(s)(k_2-k_2^{(0)}))'+B(s)(k_1-k_1^{(0)})k_3-C(s)(k_3-k_3^{(0)})+f_1=0 
\; , \label{ice} \\
(C(s)(k_3-k_3^{(0)}))'+B(s)(k_1^{(0)}k_2-k_2^{(0)}k_1)=0 \; . \label{icf} 
\end{eqnarray}
%

The components of the twist vector are expressed as:
%
\begin{eqnarray}
k_1&=&k_F(s)\sin\chi \; , \label{hchi_1} \\
k_2&=&k_F(s)\cos\chi \; , \label{hchi_2} \\
k_3&=&\chi'+\tau_F(s) \; . \label{hchi_3}
\end{eqnarray}
%

In order to obtain the simplest solution for the static Kirchhoff
equations Eqs. (\ref{ica}--\ref{icf}) with the intrinsic curvature given by
Eqs. (\ref{ks0_1}--\ref{ks0_3}) and (\ref{int_a}--\ref{int_b}) we
shall look for a solution such that $\chi=\xi$ in
Eqs. (\ref{hchi_1}--\ref{hchi_3}). This solution preserves the intrinsic twist
density of the helical structure. In this case, the Eq. (\ref{icf})
becomes simply $[C(s)(\tau_F-\tau_F^{(0)})]'=0$ or
\begin{equation}
\label{M3C}
M_3=C(s)(\tau_F-\tau_F^{(0)})=\mbox{Constant} \; , 
\end{equation}
and we obtain the following differential equations for the curvature
$k_F(s)$, and the torsion $\tau_F(s)$, of the rod:
\begin{equation}
\label{dif_ic}
\begin{array}{ll}
[M_3\,k_F-B\,\tau_F(k_F-k_F^{(0)})]'-[B(k_F-k_F^{(0)})]'\,\tau_F
=0 \; , \\

[B(k_F-k_F^{(0)})]''+\tau_F[M_3\,k_F-B\,\tau_F(k_F-k_F^{(0)})]
-f_3\,k_F=0 \; , \\

[B(k_F-k_F^{(0)})]'\,k_F+f'_3=0 \; ,   
\end{array}
\end{equation}
where we have omitted the dependence on $s$ to simplify the notation.
In order to obtain a helical solution of these equations we apply the
{\it Lancret's theorem}, Eq. (\ref{lancret}), to the
Eqs. (\ref{dif_ic}). We obtain the following results:
%
\begin{eqnarray}
&&f_3(s)=0 \; , \label{free_ic_a} \\
&&[B(s)(k_F(s)-k_F^{(0)}(s))]'=0 \;  \Rightarrow \; 
k_F(s)-k_F^{(0)}(s)=\frac{K}{B(s)} \; , \label{free_ic_b} \\
&&[B(s)(\tau_F(s)-\tau_F^{(0)}(s))]'=0 \; \Rightarrow \; 
\tau_F(s)-\tau_F^{(0)}(s)=\frac{T}{B(s)} \; , \label{free_ic_c}
\end{eqnarray}
%
where $K$ and $T$ are integration constants. From Eqs. (\ref{M3C}) and
Eq. (\ref{free_ic_c}) we obtain
\begin{equation}
\label{TT}
T=\frac{B(s)}{C(s)}M_3 \; ,
\end{equation}
so that the ratio $B(s)/C(s)$ has to be constant.

From Eqs. (\ref{free_ic_b}), (\ref{free_ic_c}), (\ref{int_a}),
(\ref{int_b}) and (\ref{lancret}) we have
\begin{equation}
\label{tanalfa_ic}
\frac{k_F(s)}{\tau_F(s)}=\frac{K+K_0}{T+T_0}=\tan{\alpha} \; .
\end{equation}
From Eqs. (\ref{icd}), (\ref{ice}), (\ref{free_ic_b}) and
(\ref{free_ic_c}) it follows that
\begin{equation}
\label{f1f2_ic}
\begin{array}{l}
f_1=0 \; ,\\
f_2=0 \; .
\end{array}
\end{equation}
Therefore, the helical solution given by Eqs. (\ref{free_ic_a}--\ref{free_ic_c})
(obtained imposing $\chi=\xi$) is a free standing helix
(${\mathbf{F}}=(f_1,f_2,f_3)=0$).

It follows from Eqs. (\ref{free_ic_a}--\ref{free_ic_c}), (\ref{int_a}) and (\ref{int_b})
that the solutions for the curvature $k_F(s)$, and the torsion
$\tau_F(s)$, of the rod with helical intrinsic curvature are similar
to those of intrinsically straight rods,
Eqs. (\ref{tf_1}--\ref{kf_1}). Therefore, rods with intrinsic
curvature and a non-constant bending coefficient given by
Eq. (\ref{r2}) (Eq. (\ref{r3})) can have a three-dimensional
configuration similar to that displayed in Fig. \ref{fig1}b
(Fig. \ref{fig1}c).

\section{Conclusions}

The existence of helical configurations for a rod with non-constant
stiffness has been investigated within the framework of the Kirchhoff
rod model. Climbing and spiralling solutions of planar rods have been
studied by Holmes {\it et. al.}~\cite{holmes2}. Here, we have shown
that helical spiralling three-dimensional structures are possible
solutions of the static Kirchhoff equations for an inhomogeneous rod.

From the static Kirchhoff equations, we derived the set of
differential equations (\ref{dif_1}--\ref{dif_3}) for the curvature
and the torsion of the centerline of a rod whose bending coefficient
is a function of the arclength $s$. We have shown that the {\it
circular helix} is the type of helical solution obtained when $B(s)$
is constant, independently of the rod being homogeneous or
inhomogeneous.

Though the differential equations for the curvature and torsion are
general, we have obtained only the simplest helical solutions
(Eqs. (\ref{sol1}) and (\ref{tf_1}--\ref{kf_1})), obtained when
the {\it Lancret's theorem} is applied to the differential
equations. We show that these solutions are free standing and that the
curvature and torsion depend directly on the form of variation of the
bending coefficient. Figures \ref{fig1}b and \ref{fig1}c are examples
of helical solutions of inhomogeneneous rods whose bending
coefficients are given by Eqs. (\ref{r2}) and (\ref{r3}). The helical
structure displayed in Fig. \ref{fig1}b is a {\it conical helix} since
the projected curve onto the plane perpendicular to the helical axis
is a {\it logarithmic spiral}, {\it i. e.}, $1/k_1(s)$ is a linear
function of $s$~\cite{dirk}.  

In the particular case of an inhomogeneous rod with the intrinsic
curvature defined by Eqs. (\ref{ks0_1}--\ref{ks0_3}) and
(\ref{int_a}--\ref{int_b}), with $B(s)/C(s)$ constant, we also obtain
the helical solutions displayed in Figs. \ref{fig1}b and
\ref{fig1}c. The tendrils of some climbing plants present a
three-dimensional structure similar to that displayed in
Fig.~\ref{fig1}b. In these plants, the cross-section diameter of the
tendrils varies along them, giving rise to non-constant bending
coefficient, and the differential growth of the tendrils produces
intrinsic curvature~\cite{tyler}. The bending and twisting
coefficients of a continuous filament with circular cross-section are
proportional to its moment of inertia $I$. It implies that $B(s)/C(s)$
is constant for an inhomogeneous rod. Therefore, the tendrils of
climbing plants can be well described by the Kirchhoff model for an
inhomogeneous rod with a linear variation of the bending stiffness.

\section*{Acknowledgements}
This work was partially supported by the Brazilian agencies FAPESP,
CNPq and CAPES. The authors would like to thank Prof. Manfredo do
Carmo for valuable informations about the Lancret's theorem.

\begin{appendix}

\section{Appendix: The differential equations for the curvature and 
torsion}

Here, we shall derive the
Eqs. (\ref{dif_1}--\ref{dif_3}). Substitution of
Eqs. (\ref{hel_k1}--\ref{hel_k3}) into Eqs. (\ref{a}--\ref{f}) gives:
%
\begin{eqnarray}
\label{es1_r}
f'_1-f_2\,(\xi'+\tau_F)+f_3\, k_F\cos\xi=0 \; , \label{ar}  \\
f'_2+f_1\,(\xi'+\tau_F)-f_3\, k_F\sin\xi=0 \; , \label{br}  \\
f'_3-f_1\, k_F\cos\xi+f_2\, k_F\sin\xi=0 \; , \label{cr}  \\
(B(s)\, k_F\sin\xi)'+(C(s)-B(s))\, k_F\cos\xi\,(\xi'+\tau_F)-f_2=0 \; , 
\label{dr} \\
(B(s)\, k_F\cos\xi)'-(C(s)-B(s))\, k_F\sin\xi\,(\xi'+\tau_F)+f_1=0 \; , 
\label{er} \\
(C(s)\ (\xi'+\tau_F))'=0 \; . \label{fr} 
\end{eqnarray}
%

First, we extract $f_1$ and $f_2$ from Eqs. (\ref{er}) and (\ref{dr}),
respectively:
%
\begin{eqnarray}
\label{fs}
&&f_1=-(B(s)\,k_F)'\cos\xi+[M_3\,k_F-B(s)\,k_F\,\tau_F]\sin\xi
\; , \label{f1r} \\
&&f_2=(B(s)\,k_F)'\sin\xi+[M_3\,k_F-B(s)\,k_F\,\tau_F]\cos\xi
\; , \label{f2r}
\end{eqnarray}
%
where $M_3=C(s)\,(\xi'+\tau_F)$ is the torsional moment of the rod
that is constant by Eq. (\ref{fr}). Differentiating $f_1$ and $f_2$
with respect to $s$, substituting in Eqs. (\ref{ar}) and (\ref{br}),
respectively, and using Eqs. (\ref{f1r}) and (\ref{f2r}), gives the
following equations:
\begin{equation}
\label{eq_A}
\begin{array}{ll}
\left\{-(B(s)\,k_F)''-\tau_F[M_3\,k_F-B(s)\,k_F\,\tau_F]
+f_3\,k_F\right\}\,\cos\xi \, \, \, + \\
\left\{[M_3\,k_F-B(s)\,k_F\,\tau_F]'
-\tau_F(B(s)\,k_F)'\right\}\, \sin\xi = 0 \; ,
\end{array}
\end{equation}
\begin{equation}
\label{eq_B}
\begin{array}{ll}
\left\{(B(s)\,k_F)''+\tau_F[M_3\,k_F-B(s)\,k_F\,\tau_F]
-f_3\,k_F\right\}\,\sin\xi \, \, \, + \\
\left\{[M_3\,k_F-B(s)\,k_F\,\tau_F]'
-\tau_F(B(s)\,k_F)'\right\}\, \cos\xi = 0 \; .
\end{array}
\end{equation}
Multiplying Eq. (\ref{eq_A}) (Eq. (\ref{eq_B})) by $\sin\xi$
($\cos\xi$) and then adding the resulting equations, we obtain the
Eq. (\ref{dif_1}) for the curvature and torsion:
\begin{equation}
\label{dif1r}
[M_3\,k_F-B\,k_F\,\tau_F]'-(B\,k_F)'\tau_F=0 \; .
\end{equation}
Multiplying Eq. (\ref{eq_A}) (Eq. (\ref{eq_B})) by $-\cos\xi$
($+\sin\xi$) and then adding the resulting equations, we obtain the
Eq. (\ref{dif_2}):
\begin{equation}
\label{dif2r}
(B\,k_F)''+k_F\,\tau_F(M_3-B\tau_F)-f_3\,k_F=0 \; .
\end{equation}
Finally, the Eq. (\ref{dif_3}) is obtained by substituting
Eqs. (\ref{f1r}) and (\ref{f2r}) in Eq. (\ref{cr}):
\begin{equation}
\label{dif3r}
(B\,k_F)'+f_3'=0 \; .
\end{equation}

\end{appendix}


\newpage  

\begin{thebibliography}{99}
\bibitem{whip} A whip is a nonhomogenous thread with varying radius of
cross section.
\bibitem{foot2} The references in \cite{dill} can be seen for a 
complete derivation of the Kirchhoff equations.
\end{thebibliography}
\end{document}